# Approximation algorithms for two-state anti-ferromagnetic spin systems on bounded degree graphs

Alistair Sinclair[*]    Piyush Srivastava[†]    Marc Thurley[‡]

**Abstract**

In a seminal paper [12], Weitz gave a deterministic fully polynomial approximation scheme for counting exponentially weighted independent sets (equivalently, approximating the partition function of the hard-core model from statistical physics) on graphs of degree at most $d$, up to the critical activity for the uniqueness of the Gibbs measure on the infinite $d$-regular tree. More recently Sly [10] showed that this is optimal in the sense that if there is an FPRAS for the hard-core partition function on graphs of maximum degree $d$ for activities larger than the critical activity on the infinite $d$-regular tree then NP = RP. In this paper, we extend Weitz's approach to derive a deterministic fully polynomial approximation scheme for the partition function of the anti-ferromagnetic Ising model with arbitrary field on graphs of maximum degree $d$, up to the corresponding critical point on the $d$-regular tree. The main ingredient of our result is a proof that for two-state anti-ferromagnetic spin systems on the $d$-regular tree, weak spatial mixing implies strong spatial mixing. This in turn uses a message-decay argument which extends a similar approach proposed recently for the hard-core model by Restrepo *et al* [9] to the case of the anti-ferromagnetic Ising model with arbitrary field. By a standard correspondence, these results translate to arbitrary two-state anti-ferromagnetic spin systems with soft constraints.

## 1 Introduction

### 1.1 Background

Spin systems are a general framework for modeling nearest-neighbor interactions on graphs, and are widely studied in both statistical physics and applied probability. A spin system consists of a large collection of nodes, each of which may be in one of a fixed number of states called *spins*. A neighborhood structure is specified by edges between the nodes. Interactions between neighboring nodes are determined by *edge potentials*, which assign an energy value to each edge based on the spin values of its endpoints. In addition, there are *vertex potentials* which assign an energy value to each node based on the value of its spin. For any configuration $\sigma$ of spins on the nodes, the energy $H(\sigma)$ is just the sum of its edge and vertex energies. Based on the Gibbs formalism from statistical physics, the probability of finding the system in configuration $\sigma$ is then proportional to the weight $w(\sigma) = \exp(-H(\sigma))$.

In this paper, we concentrate on *two-state* spin systems, where each vertex can be in one of two states, referred to as "$+$" and "$-$". Such a system can be defined by specifying a $(+,+)$ *edge activity* $\beta$, a $(-,-)$ edge activity $\gamma$, and a *vertex activity* $\lambda$, where $\beta$, $\gamma$ and $\lambda$ are non-negative parameters. For a graph $G = (V, E)$, a *configuration* $\sigma : V \mapsto \{+,-\}$ is an assignment of $+$ and $-$ spins to the vertices of $G$. The weight $w(\sigma)$ of the configuration $\sigma$ is given by

$$w(\sigma) = \lambda^{m(\sigma)} \beta^{n_+(\sigma)} \gamma^{n_-(\sigma)}, \tag{1}$$

---

[*]Computer Science Division, UC Berkeley. Email: sinclair@cs.berkeley.edu. Supported in part by NSF grant CCF-1016896.

[†]Computer Science Division, UC Berkeley. Email: piyushsr@eecs.berkeley.edu. Supported by the Berkeley Fellowship for Graduate Study and NSF grant CCF-1016896. Part of this work was done when this author was a research intern at Microsoft Research India.

[‡]Centre de Recerca Matemàtica, Bellaterra. Email: marc.thurley@googlemail.com. Supported in part by a postdoctoral fellowship of the German Academic Exchange Service (DAAD) and by Marie Curie Intra-European Fellowship 271959. Part of this work was done while this author was a postdoctoral scholar at UC Berkeley.

where $m(\sigma)$ denotes the number of vertices assigned spin $-$, and $n_+(\sigma)$ (respectively, $n_-(\sigma)$) denotes the number of edges for which both endpoints are assigned spin $+$ (respectively, $-$). The *partition function* of the model is defined as

$$Z = \sum_{\sigma \in \{+,-\}^V} w(\sigma). \tag{2}$$

The partition function, in addition to being a natural weighted generalization of the notion of counting, is a fundamental quantity in statistical physics. For example, it is the normalizing factor in the Gibbs distribution: the probability of occurrence of configuration $\sigma$ is given by $\mu(\sigma) = w(\sigma)/Z$. In addition, many other properties of the model can be deduced by studying the partition function [3].

As a simple concrete example of a two-state spin system, consider the setting $\beta = 1$ and $\gamma = 0$ (so that configurations with adjacent "$-$" spins are assigned weight zero, and thus prohibited), known as the *hard-core* model. The associated Gibbs distribution is a weighted measure on independent sets in the graph $G$, in which any independent set $U$ has weight $\lambda^{|U|}$. Another important class of examples, known as the *Ising model*[1], is obtained by setting $\beta = \gamma > 0$. There is a significant qualitative difference between the Ising model with $\beta = \gamma > 1$ (the *ferromagnetic* case) and with $\beta = \gamma < 1$ (the *anti-ferromagnetic* case). The latter is an example of a "repulsive" model, which means that the edge potentials assign higher weights to edges with different spins at their endpoints, while the ferromagnetic case is "attractive" (higher weights are assigned to edges with the same spin at their endpoints). The parameter $\lambda$ can be identified with an "external field", i.e., a bias associated with each spin. The case $\lambda = 1$ corresponds to zero field, while $\lambda < 1$ and $\lambda > 1$ correspond to positive and negative fields respectively. More generally, we will refer to any two-state system satisfying $\beta\gamma > 1$ as *ferromagnetic*, and any satisfying $\beta\gamma < 1$ as *anti-ferromagnetic*. Also, a model satisfying $\beta\gamma > 0$ is said to have *soft constraints* (in the sense that no combination of spin values at adjacent vertices is prohibited). In a sense to be made precise later (see Appendix A), Ising models capture arbitrary two-state spin systems with soft constraints; in particular, the two descriptions are equivalent on regular graphs. For this reason we will henceforth focus mainly on Ising models.

The theory of spin systems derives in large part from considering the limiting behavior of the Gibbs distribution as the size of the underlying graph goes to infinity. Based on the above formalism for finite graphs, one may define a Gibbs measure $\mu$ on an infinite graph $\mathcal{G}$ by requiring that the marginal distribution on any finite subgraph $\mathcal{H}$, conditional on the configuration on $\mathcal{G}\backslash\mathcal{H}$, is given by equation (1). (Here the spins in $\mathcal{G}\backslash\mathcal{H}$ act as a fixed *boundary condition* in (1).) It is a well known result in the statistical physics literature (see, for example, [3]) that at least one such measure $\mu$ can always be defined. However, for certain values of the parameters of the spin system there may be multiple solutions for $\mu$, in which case the Gibbs measure is said to be *non-unique*.

We will now look at the phenomenon of non-uniqueness more closely in the special case when the infinite graph $\mathcal{G}$ is a $d$-ary tree.[2] As noted above, the anti-ferromagnetic Ising model captures all two-state anti-ferromagnetic spin systems with soft constraints on regular graphs, and hence it is sufficient to consider the Ising case. Consider an anti-ferromagnetic Ising model on the $d$-ary tree with edge activity $\beta(=\gamma)$ and vertex activity $\lambda$. It turns out that if $\beta \geq \frac{d-1}{d+1}$ then the Gibbs measure is unique for all values of $\lambda$. In particular, in the zero-field case $\lambda = 1$, the Gibbs measure is unique *if and only if* $\beta \geq \frac{d-1}{d+1}$. However, when $\beta < \frac{d-1}{d+1}$, the Gibbs measure is no longer unique for all values of the vertex activity $\lambda$. In this case, there exists a *critical activity* $\lambda_c(\beta, d) \geq 1$ such that the Gibbs measure is unique if and only if $|\log \lambda| \geq \log \lambda_c(\beta, d)$. We sketch the curves of $\log \lambda_c(\beta, d)$ in Figure 1, for $d = 5$ and $d = 13$. The area below the curves is the non-uniqueness region. We note here that the non-uniqueness region is monotonically increasing with degree, so the curve for $d = 5$ lies strictly below that for $d = 13$. Also, note that the curves intersect the $\beta$-axis at $\beta = \frac{d-1}{d+1}$.

---

[1] The description of the Ising model given here differs slightly from the more popular description in terms of edge and vertex potentials outlined in the first paragraph. However, translating between the two descriptions is easy; see Appendix A.

[2] We remark here that the infinite $(d+1)$-regular tree and the infinite $d$-ary tree show exactly the same behavior with respect to the uniqueness of the Gibbs measure. This follows immediately from the fact that the $(d+1)$-regular tree can be viewed as a root attached to the roots of $d+1$ infinite $d$-ary trees. We shall thus move freely between these two objects for ease of exposition throughout the paper.

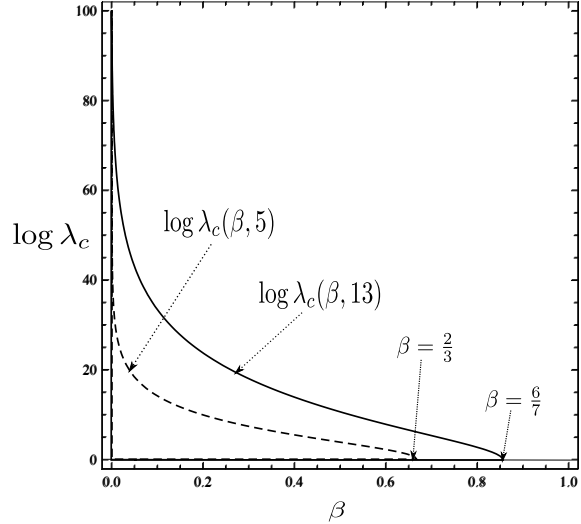

Figure 1: $\log \lambda_c(\beta, d)$ for $d = 5$ and $d = 13$. The curves intersect the $\beta$-axis at $\beta = \frac{2}{3}$ and $\beta = \frac{6}{7}$ respectively.

The phenomenon of non-uniqueness of the Gibbs measure can also be described in terms of the more algorithmic notion of *decay of correlations*. We stick to our example of the infinite $d$-ary tree. Fix a vertex $v$ in the tree, and let $S_l$ be the set of vertices in the tree at distance at least $l$ from $v$. Let $q_v(l, \sigma)$ be the probability of having spin $+$ at $v$ conditional on the configuration on $S_l$ being $\sigma$. It turns out that uniqueness of the Gibbs measure is equivalent to the condition that the inequality

$$|q_v(l, \sigma) - q_v(l, \tau)| \leq \exp(-\Omega(l)) \tag{3}$$

holds for any two configurations $\sigma$ and $\tau$ on $S_l$.[3] The above condition is referred to in the literature as *weak spatial mixing*.

It has been believed for a long time (and proved in various manifestations) that there is an intimate relationship between weak spatial mixing and the running time of algorithms for approximating the associated partition function: roughly speaking, in the uniqueness region (where there is decay of correlations), the system should be amenable to local algorithms and thus be computationally tractable. A spectacular result in this direction was Weitz's fully polynomial deterministic approximation scheme (FPTAS) for the partition function of the hard-core model, which works on all graphs of degree at most $d + 1$ for all activities $\lambda$ less than the critical activity $\lambda_c(d)$ for the uniqueness of the Gibbs measure on the infinite $d$-ary tree [12]. This is even more interesting in light of a recent breakthrough due to Sly [10] (see also [2]), who showed that the existence of an FPRAS for the partition function of the hard-core model on graphs of degree at most $d + 1$ for activities larger than $\lambda_c(d)$ would imply that NP = RP. Thus the range of validity of Weitz's algorithm is optimal.

Weitz [12] gave a general two-step framework for designing deterministic algorithms for approximating partition functions of two-state spin systems. To describe this framework, we begin with the standard observation that in order to get an FPTAS for the partition function, it is sufficient to give an FPTAS for the probability of having spin $+$ at any given vertex $v$. The first component of the framework is a combinatorial reduction, which shows that the problem of approximating this probability for a general two-state spin system on a graph $G$ of maximum degree $d + 1$ can be reduced to the problem of approximating the same probability on a related finite subtree of the infinite $(d+1)$-regular tree rooted at $v$, in which the spins of some of the vertices are fixed to certain values (this is the so-called *self-avoiding walk tree* of the graph $G$). We emphasize that this is a model-independent reduction, and depends only upon the fact that the number of spin values is two. The associated self-avoiding walk tree, however, may be exponential in the size of the original graph $G$, and thus one needs to show that it is sufficient to truncate the tree at a depth logarithmic in the size of $G$ in order to obtain a good approximation. However, since some of the fixed vertices in the tree might be very close to the root $v$, it is not possible to argue using weak spatial mixing that a logarithmic depth of recursion suffices for approximating the partition function (because the parameter $l$ in equation (3) must be taken to be the minimum distance of a fixed vertex from the root).

---

[3]To be precise, this condition does not hold on the boundary of the uniqueness region, that is, for $|\log \lambda| = \log \lambda_c(\beta, d)$; at this critical value, the l.h.s of equation (3) still decays to 0 with $l$, but not at an exponential rate. We will focus on the interior of this region, and by a slight abuse of terminology refer to it as the "uniqueness region".

Accordingly, the second component of Weitz's framework is to establish that, for the spin system in question, weak spatial mixing on the infinite $d$-ary tree is in fact equivalent to *strong spatial mixing*, which roughly states that the exponential decay of point-to-set correlations (3) guaranteed by weak spatial mixing holds also when the spins at an arbitrary set of vertices are fixed to arbitrarily chosen values (see Section 2 for a precise definition). Weitz [12] established this fact for the hard core model, using a step-by-step comparison of ratios of occupation probabilities on the standard $d$-ary tree and on the modified tree with fixed vertices. It was claimed in [12] that such a result holds also for the anti-ferromagnetic Ising model, but to the best of our knowledge no proof of this fact (except in the special zero-field case where $\lambda = 1$; see [9, 13]) has so far been published.

## 1.2 Contributions

In this paper, we give a proof of the fact that for the anti-ferromagnetic Ising model with any field, weak spatial mixing implies strong spatial mixing on the $d$-ary tree. Formally, we have the following theorem.

**Theorem 1.1** *For the anti-ferromagnetic Ising model with arbitrary field on the $d$-ary tree with $d \geq 2$, weak spatial mixing implies strong spatial mixing.*

Notice that it is easy to see that this holds also for the infinite $(d+1)$-regular tree, since the $(d+1)$-regular tree and the $d$-ary tree differ only in the degree of the root. We also note that by the translation described in Appendix A, Theorem 1.1 holds also for arbitrary anti-ferromagnetic two-state spin systems with soft constraints.

Given Weitz's general reduction described above, we obtain as an almost immediate consequence of Theorem 1.1 an FPTAS for the partition function of the anti-ferromagnetic Ising model on graphs of maximum degree at most $d+1$ throughout the uniqueness region of the Gibbs measure on the $d$-ary tree.

**Corollary 1.2** *Let $d \geq 2$. Consider an anti-ferromagnetic Ising model with parameters $\beta$ and $\lambda$. For $\beta$ and $\lambda$ in the interior of the uniqueness region of the $d$-ary tree, there is a deterministic polynomial time approximation scheme for the partition function of the associated spin system on graphs of degree at most $d+1$.*

By the translation described in Appendix A, we can extend this result to general two-state anti-ferromagnetic spin systems. The difference is that the critical activity may now differ for vertices of different degrees. Let $\lambda_c(\beta, d)$ be the critical activity for the anti-ferromagnetic Ising model described above (and defined formally in Section 2.2.1). Then, we have the following corollary.

**Corollary 1.3** *Let $d \geq 2$. Consider an anti-ferromagnetic two-state spin system with parameters $\beta$, $\gamma$ and $\lambda$. Let $\beta'$ be the edge potential for the equivalent anti-ferromagnetic Ising model. Let $\mathcal{G}$ be the class of graphs with maximum degree $d+1$ in which every vertex $v$ satisfies the condition*

$$|\log \lambda_v| \triangleq |\log \lambda + \frac{\deg(v)}{2}(\log \gamma - \log \beta)| > \log \lambda_c(\beta', d).$$

*Then there is a deterministic polynomial time approximation scheme for the partition function of the associated spin system on graphs in the class $\mathcal{G}$. In particular, the class $\mathcal{G}$ includes all $(d+1)$-regular graphs when $\beta, \gamma$ and $\lambda$ are in the interior of the uniqueness region of the $d$-ary tree.*

We briefly sketch the approach we use to prove our main technical result, namely that weak spatial mixing implies strong spatial mixing (Theorem 1.1). Inspired by recent work of Restrepo *et al* [9], we design a "message" (i.e., an invertible function of the probability of a vertex having spin + ) such that "disagreements" in the message decay by a constant factor at each vertex of the tree. The challenge is to ensure that such a message can be designed for all points in the uniqueness region of the $d$-ary tree. For the special zero-field case of the anti-ferromagnetic Ising model (when $\lambda = 1$), such a message is well known [13]. However, this message does not work up to the threshold for general vertex potentials $\lambda$. Restrepo *et al* [9] recently derived a message which works up to the tree uniqueness threshold for the hard-core model. For the general anti-ferromagnetic Ising model, such a message turns out to be more complex than those known for the zero-field case and for the hard-core model. Our message is defined at the beginning of Section 3, and the requisite decay property is established in Section 4.

We conjecture that our proof of strong spatial mixing based on stepwise decay of messages may lead to further consequences. For example, as shown by Restrepo *et al* [9], the message decay property can be used to extend Weitz's algorithm by exploiting the structure of special classes of graphs to obtain approximation algorithms beyond the tree threshold for those graphs. In addition, our proof demonstrates the versatility of the message approach.

Finally, we point out a byproduct of our approach that may be of independent interest. The exact value of the critical field $\lambda_c(\beta, d)$ as a function of $(\beta, d)$ is apparently widely accepted folklore, but the only derivation we could find for it in the literature [3, p. 255] does not provide a formal proof, appealing instead to numerical evidence. We point out that our proof of strong spatial mixing does not assume knowledge of $\lambda_c(\beta, d)$, and in fact derives its value as a byproduct (see the Remark following Theorem 2.5 for more details). Thus our approach gives a proof of the location of the uniqueness threshold $\lambda_c(\beta, d)$ of two-state anti-ferromagnetic spin systems.

**Remark:** After obtaining our message-decay proof, we received a sketch of Weitz's original unpublished proof [11]. It is interesting to note that that proof is quite different from ours, and employs a delicate two-step analysis of the tree recursion described in Section 2. For reasons mentioned above, we believe that our message-decay proof, in addition to being the first published version of this result, is potentially more robust and flexible than Weitz's approach; for example, it is not clear how to adapt Weitz's analysis to obtain stronger results for special classes of graphs such as lattices, as is done in [9].

## 1.3 Related work

Our work is mainly motivated by the deterministic counting algorithm of Weitz [12], which was the first to show an interesting connection between the running time of an algorithm not related to Markov chain Monte Carlo and the phase transition phenomenon for spin systems. On the complexity side, using a randomized gadget first proposed by Dyer, Frieze and Jerrum [1] and analysed further by Mossel, Weitz and Wormald [8], Sly [10] proved that if there is an FPRAS for the partition function of the hard-core model on graphs of degree at most $d$ in the non-uniqueness region of the $d$-regular tree, then NP = RP, thus showing that the range of validity of Weitz's algorithm is optimal. Technically Sly's result holds only sufficiently close to the boundary of the uniqueness region; this restriction was mostly removed in a recent paper of Galanis *et al* [2]. For the case of unbounded degree graphs, Goldberg, Jerrum and Paterson [5] showed that approximating the partition function for the zero-field case ($\lambda = 1$) is NP-hard in the interior of the square $0 \leq \beta, \gamma \leq 1$.

A related problem is to get exponential lower bounds on the mixing time of any local Markov chain (Glauber dynamics) that samples from the hard-core and anti-ferromagnetic Ising models. Mossel, Weitz and Wormald [8] and Gerschenfeld and Montanari [4] showed that beyond the uniqueness threshold for $d$-regular trees, Glauber dynamics for these models can take exponential time to mix on $d$-regular graphs. Gerschenfeld and Montanari also show that for these models on random regular graphs, this threshold for slow mixing is also the threshold beyond which the reconstruction problem is solvable, and pointed out that these results therefore establish the existence of $d$-regular graphs on which the reconstruction problem is solvable beyond the uniqueness threshold for the $d$-regular tree [4].

On the algorithmic side, an analysis of Weitz's algorithm for the zero-field case of the anti-ferromagnetic

Ising model appears in [13]. There has been some subsequent progress on the hard-core model on special classes of graphs too: recently, Restrepo *et al* [9] used a message-decay proof to get improved strong spatial mixing thresholds on the 2D integer lattice for the hard core model. They achieved this by exploiting the special structure of self-avoiding walk trees obtained when Weitz's reduction is applied to the lattice. The message-decay proof turns out to be crucial in tightening the analysis to obtain strong spatial mixing over a wider range of parameters for these special trees. Much more is known about algorithms for the ferromagnetic case: Jerrum and Sinclair [6] gave an FPRAS for the Ising model with arbitrary field on graphs of arbitrary degree, while Goldberg, Jerrum and Paterson [5] showed how to extend this to the whole of the ferromagnetic region $\beta\gamma > 1$ with $\lambda = 1$. The latter paper [5] also gave an FPRAS for the partition function on graphs of arbitrary degree for parts of the anti-ferromagnetic region $\beta\gamma < 1$. However, the results of [5] when restricted to bounded degree graphs do not hold throughout the uniqueness region and hence are incomparable to ours.

In a recent paper, Li, Lu and Yin [7] improve upon the work of [5]. They consider two-state anti-ferromagnetic spin systems with zero field ($\lambda = 1$) on general graphs, and derive a condition under which an FPTAS exists for approximating the partition function. Their condition requires that $(\beta, \gamma)$ lies in the *intersection* of the uniqueness regions for *all* possible degrees $d$. (Note that in the $(\beta, \gamma)$ parameterization, this is a non-trivial region of the $\beta\gamma$-plane.) However, for any fixed degree $d$, this region is smaller than the uniqueness region for $d$, which is the range of validity obtained in our present paper. We also point out an important qualitative difference between the parameterization we use (edge potential $\beta$ and vertex field $\lambda$) and the parameterization via two edge potentials ($\beta$ and $\gamma$) used by Li *et al*: while the $(\beta, \gamma)$ parametrization is *not* monotonic, in the sense that uniqueness on a $d$-regular tree does not imply uniqueness on a $(d-1)$-regular tree, the $(\beta, \lambda)$ parametrization *is* monotonic. In fact, as our results show, uniqueness on the $d$-regular tree implies strong spatial mixing, and hence uniqueness, on all $d'$-regular trees for $d' \leq d$ in the $(\beta, \lambda)$ parameterization.

## 2 Preliminaries

### 2.1 Notation

We will mostly follow the notational conventions of [5]. Given a graph $G = (V, E)$, a two-state spin configuration is defined as an assignment $\sigma : V \mapsto \{+, -\}$ of spins to the vertices. Weights for different configurations are computed in terms of the $(+, +)$-edge activity $\beta$, the $(-, -)$ edge activity $\gamma$ and a vertex activity $\lambda$, and are given by

$$w(\sigma) = \lambda^{m(\sigma)} \beta^{n_+(\sigma)} \gamma^{n_-(\sigma)}, \qquad (4)$$

where given the configuration $\sigma$, $m(\sigma)$ denotes the number of vertices assigned spin $-$, and $n_+(\sigma)$ (respectively, $n_-(\sigma)$) denotes the number of edges for which both endpoints are assigned spin $+$ (respectively, $-$). The *partition function* is defined as

$$Z = \sum_{\sigma \in \{-1,1\}^V} w(\sigma).$$

We remark that this representation can be easily translated to the usual description in terms of edge potentials and vertex field: for completeness we give the translation in Appendix A.

**Definition 2.1 (Occupation probability).** *Given a vertex $v$ in the graph $G$, the* occupation probability *$p_v$ is the probability that $v$ is assigned spin $+$ in a random configuration $\sigma$ sampled according to the weights defined in equation (4).*

### 2.2 The Ising model

The Ising model corresponds to the case $\beta = \gamma$. The model is *ferromagnetic* when $\beta > 1$ and *anti-ferromagnetic* when $\beta < 1$ (the case $\beta = 1$ is trivial). The zero-field case corresponds to $\lambda = 1$, the positive field case to $\lambda < 1$ and the negative field case to $\lambda > 1$. As shown in Appendix A, on $d$-regular graphs the Ising model is equivalent to general two-state spin systems. Thus, in the rest of this paper, we will concentrate mostly on the Ising case. On non-regular graphs the equivalence still holds; however, the vertex activity $\lambda$ in the Ising model may then be different on different vertices. The adaptation of our results to this setting is described in Corollary 1.3.

### 2.2.1 Phase transition.

The anti-ferromagnetic Ising model exhibits a uniqueness phase transition on the $d$-ary tree for $d \geq 2$. In particular, one can define a critical activity $\lambda_c(\beta, d)$ as follows.

**Definition 2.2 (Critical activity).** *Consider the anti-ferromagnetic Ising model on an infinite $d$-ary tree with edge activity $\beta$ and vertex activity $\lambda$. If $\beta \geq \frac{d-1}{d+1}$ then the Gibbs measure is unique for all values of $\lambda$. If $\beta < \frac{d-1}{d+1}$, then there exists a critical activity $\lambda_c(\beta, d) \geq 1$ such that the Gibbs measure is unique if and only if $|\log \lambda| \geq \log \lambda_c(\beta, d)$.*

A consequence of uniqueness[4] of the Gibbs measure is *weak spatial mixing*, which captures a weak notion of decay of point to set correlations. Let $p_\rho(\sigma, S)$ be the probability of occupation of the root $\rho$ of an infinite $d$-ary tree when the spins of a set $S$ of nodes are fixed according to the configuration $\sigma$. Let $\delta(\rho, S)$ denote the distance of $\rho$ from the set $S$.

**Definition 2.3 (Weak spatial mixing).** *Given any two-state spin system,* weak spatial mixing *is said to hold if for any set $S$ whose distance $\delta(\rho, S)$ from the root $\rho$ of the tree is finite, and any two configurations $\sigma_1$ and $\sigma_2$, we have*

$$|p_\rho(\sigma_1, S) - p_\rho(\sigma_2, S)| \leq \exp(-\Omega(\delta(\rho, S))).$$

Notice that weak spatial mixing does not guarantee exponential decay of correlations when the set $S$ contains vertices which are very close to the root $\rho$, even when $\sigma_1$ and $\sigma_2$ differ only on vertices which are very far away from $\rho$. A related but, as the name suggests, stronger notion is that of *strong spatial mixing*, which captures the idea that fixing vertices near the root to the *same* spin should not affect the exponential decay of point-to-set correlations. We note that strong spatial mixing is not in general implied by weak spatial mixing for arbitrary spin systems; see Appendix B for a counterexample involving the ferromagnetic Ising model with appropriate parameters.

**Definition 2.4 (Strong spatial mixing).** *Given any two-state spin system,* strong spatial mixing *is said to hold if for any set $S$ whose distance $\delta(\rho, S)$ from the root $\rho$ of the tree is finite, and any two configurations $\sigma_1$ and $\sigma_2$ which differ only on a set $T \subseteq S$ of vertices, we have*

$$|p_\rho(\sigma_1, S) - p_\rho(\sigma_2, S)| \leq \exp(-\Omega(\delta(\rho, T))).$$

### 2.2.2 Phase transition and tree recursions.

It is well known (see, for example, [3]) that the uniqueness condition for two-state spin systems on $d$-ary trees can be written in terms of the number of fixed points of the recursion for occupation probabilities. Consider a subtree rooted at a vertex $v$ in the $d$-ary tree, and let $v_i, i = 1, 2, ...d$ be its children. Let $p_v$ be the occupation probability at vertex $v$ and define $R_v = \frac{1-p_v}{p_v}$. One can then write the following recurrence for $R_v$:

$$R_v = \lambda \prod_{i=1}^{d} \left( \frac{\beta R_{v_i} + 1}{\beta + R_{v_i}} \right).$$

This can easily be converted to a recurrence for occupation probabilities. Define

$$h(x) \triangleq \frac{\beta + (1-\beta)x}{1 - (1-\beta)x}.$$

---
[4]As stated in the introduction, we exclude the boundary of the uniqueness region here.

We can then write the recurrence as

$$p_v = F(p_{v_1}, p_{v_2}, \ldots, p_{v_d}) \triangleq \frac{1}{1 + \lambda \prod_{i=1}^{d} h(p_{v_i})}. \tag{5}$$

We will find it useful in what follows to consider the tree recurrence with the special boundary condition in which all vertices at some distance $l$ from the root are fixed to the same spin. In this case, by symmetry, the tree recurrence outputs the same value $p_v$ at all vertices $v$ which are at the same distance from the root. Thus, the recurrence can be simplified to a one-parameter recurrence as follows:

$$p_v = f(p_{v_1}) \triangleq \frac{1}{1 + \lambda h(p_{v_1})^d}.$$

Note that in the anti-ferromagnetic case, $h$ is an increasing function, and hence $F$ and $f$ are decreasing in each of their arguments. We also note that since $f$ is strictly decreasing in $[0,1]$, it has a unique fixed point.

In terms of the recurrence function $f$, the condition for uniqueness can be stated as follows.

**Theorem 2.5 ([3])** *For given values of $\beta$ and $\lambda$, the infinite d-ary tree has a unique Gibbs measure if and only if the two-step recurrence function $f \circ f$ has a unique fixed point. In particular, if the Gibbs measure is unique, and $(\beta, \lambda)$ are not on the boundary of the uniqueness region, then the unique fixed point $x^\star$ of $f$ satisfies*

$$f'(x^\star) > -1. \tag{6}$$

**Remark:** In [3], it is claimed (implicitly) on the basis of numerical simulations that the condition (6) is also *sufficient* for uniqueness. To be precise, the expression for the critical activity $\lambda_c(\beta, d)$ given in [3, p. 255] is exactly the same as that obtained by assuming that (6) is also a sufficient condition for uniqueness. While we believe this fact to be folklore, we have not been able to find a rigorous proof of it in the literature. With a slight abuse of terminology, we will henceforth refer to the set of $(\beta, \lambda)$ for which the fixed point $x^\star$ satisfies $f'(x^\star) > -1$ as the "uniqueness region". We will justify this terminology later (see the Remark following the proof of Theorem 1.1 in Section 4) by proving that condition (6) does indeed imply uniqueness. Thus we will obtain a rigorous proof of the expression for the critical activity appearing in [3].

### 2.3 Messages

**Definition 2.6 (Message).** *A message is a continuously differentiable function $\phi : [0,1] \mapsto \mathbb{R}$ with positive derivative.*

Note that a message is strictly increasing and hence invertible on its range. Moreover, the inverse function $\phi^{-1}$ is also a continuously differentiable function with positive derivative.

Given a recurrence function $f : [0,1] \mapsto \mathbb{R}^+$, and a message $\phi$, we denote by $f^\phi$ the function $\phi \circ f \circ \phi^{-1}$. The function $f^\phi$, which will play a crucial role in this paper, describes the evolution of the message $\phi$ under the recurrence, in the sense that $f^\phi(\phi(x)) = \phi(f(x))$. We will also need the following fact.

**Fact 2.7** *For any message $\phi$, the parameters $(\beta, \lambda)$ are in the uniqueness region if and only if $f^{\phi'}(p^\star) > -1$ at the unique fixed point $p^\star$ of $f^\phi$.*

*Proof.* Notice that since $\phi$ is strictly increasing, and $f$ has a unique fixed point $x^\star$, $f^\phi$ also has a unique fixed point $p^\star = \phi(x^\star)$. Now, we notice that $f^{\phi'}(p^\star) = f'(x^\star)$, because

$$\begin{aligned} f^{\phi'}(p^\star) &= \phi'(f(\phi^{-1}(p^\star)))f'(\phi^{-1}(p^\star))\phi^{-1'}(p^\star) \\ &= \phi'(f(x^\star))f'(x^\star)\frac{1}{\phi'(\phi^{-1}(p^\star))} \\ &= \frac{\phi'(x^\star)}{\phi'(x^\star)}f'(x^\star) \\ &= f'(x^\star), \end{aligned}$$

where in the second line we used the facts that $\phi^{-1}(p^\star) = x^\star$ and $\phi^{-1'}(y) = \frac{1}{\phi'(\phi^{-1}(y))}$, and in the third line the fact that $f(x^\star) = x^\star$. Thus, $(\beta, \lambda)$ are in the uniqueness region (as defined in the Remark following Theorem 2.5) if and only if $f^{\phi'}(p^\star) = f'(x^\star) > -1$. ∎

## 2.4 Weitz's tree reduction

As indicated in the introduction, Weitz [12] proved the following combinatorial reduction.

**Theorem 2.8** *For any two-state spin system, strong spatial mixing on the $d$-ary tree implies that there exists a deterministic fully polynomial approximation scheme for the partition function of the spin system on graphs of degree at most $d + 1$.*

## 3 Messages and contraction on the $d$-ary tree

In this section, we will prove the main technical ingredient of our result, which is expressed in the following theorem.

**Theorem 3.1** *Given $d, \beta$ and $\lambda$, there exists a message $\phi$ and a constant $c < 1$, such that the tree recurrence $g \triangleq f^\phi$ for the quantity $\phi(p_v)$ satisfies $\|g'\|_\infty \leq c < 1$, whenever $(\beta, \lambda)$ is in the uniqueness region for the $d$-ary tree.*

The above theorem says that in the uniqueness region, the single-parameter recurrence $g = f^\phi$ for the message $\phi(p_v)$ contracts at every step. (Without the message, the function $f$ itself is not contractive.) This stepwise contraction is easily seen by standard arguments to imply *weak* spatial mixing; for completeness, we give a proof in Appendix C.1. To extend the argument to *strong* spatial mixing, as required for Theorem 1.1, we need to consider a multi-parameter (vectorized) version of the message recurrence $g$, since under arbitrary boundary conditions the occupation probabilities need not be uniform. We will show in Section 4 that for our message $\phi$ in Theorem 3.1, the analysis of the vectorized version can in fact be reduced to an application of Theorem 3.1.

**Remark:** For ease of notation, in the rest of the paper we will prove our results in terms of the uniqueness threshold of the $d$-ary tree, relating it to algorithms on graphs of degree at most $d+1$. As already noted, the uniqueness thresholds on the $(d+1)$-regular tree and the $d$-ary tree coincide, and hence our results apply equally to the infinite $(d+1)$-regular tree.

We begin by setting up some notation for the proof of Theorem 3.1. Notice that in the light of Fact 2.7 the main technical challenge is to come up with a message $\phi$ such that the quantity $\left|f^{\phi'}\right|$ is maximized at the unique fixed point of $f^\phi$. Let us fix constants

$$A = d(1 - \beta^2) + (1 - \beta)^2 \text{ and } D = \frac{\sqrt{A + 4\beta} - \sqrt{A}}{2\sqrt{A}}.$$

Define

$$\phi(x) = \log\left(\frac{x + D}{1 - x + D}\right).$$

Notice that $D > 0$, so $\phi$ is a continuously differentiable function with positive derivative on the interval $[0, 1]$. Using this message we are able to prove the following.

**Lemma 3.2** *Consider the anti-ferromagnetic Ising model on a $d$-ary tree with edge activity $\beta$ and vertex activity $\lambda$. Then, defining $g = f^\phi$, $\psi = \phi^{-1}$, $\alpha = \psi(x)$ and $\eta = f(\alpha)$, we have*

$$g''(x) = (\eta - \alpha)g'(x)\psi'(x)\frac{d\beta(1 - \beta^2)(2\beta + A\alpha\eta + A(1 - \alpha)(1 - \eta))}{(\beta + \alpha(1 - \alpha)(1 - \beta)^2)(\beta + A\eta(1 - \eta))(\beta + A\alpha(1 - \alpha))}. \tag{7}$$

The proof of Lemma 3.2 is somewhat technical and is given in Appendix D.

Before proceeding with the technical development, we pause to give some comments on the design of our message. Notice that the requirement that the derivative of the function $g = f^\phi$ should have its maximum magnitude at the unique fixed point of $g$ does not immediately lead to a solution for $\phi$, and thus we must resort to some educated guesswork for the functional form of $\phi$. Our choice is guided by the intuition that, by analogy with the zero field case, where it is well known that the simple message $\phi(x) = \log\left(\frac{x}{1-x}\right)$ is sufficient, a log ratio of probabilities shifted by an additive constant $D$ to account for the field should be appropriate. The choice of $D$ is then determined by the above requirement. An important additional property of our message is that, perhaps surprisingly, it does not depend upon the vertex potential $\lambda$, but only upon the edge potential $\beta$ and the degree $d$; this is reflected in the fact that the additive shift $D$ is the same for both probabilities. This property will be important in extending our algorithm to the setting of general two-state anti-ferromagnetic spin systems in Corollary 1.3.

**Lemma 3.3** *Let $g = f^\phi$, with the message $\phi$ defined above. Then $|g'(x)|$ is maximized at the unique positive fixed point of $g$.*

*Proof.* We use the notation established in Lemma 3.2, where we derived an expression for $g''(x)$ in equation (7). It is easy to see that ignoring the factor $(\eta - \alpha)$, the rest of the right hand side of equation (7) is negative: this is because $g$ is a decreasing function, while $\psi$, being the inverse of the increasing function $\phi$, is increasing. Also, we have $0 < \beta < 1$ (in the anti-ferromagnetic case) and $0 \leq \alpha, \eta \leq 1$ (since they are probabilities), so that the fractions appearing on the right hand side are positive.

Let $x^\star$ be the unique fixed point of the strictly decreasing function $g$. From the above discussion, it follows that the sign of $g''(x)$ is the opposite of the sign of $\eta - \alpha = f(\psi(x)) - \psi(x)$. Notice that $\eta - \alpha$ is strictly positive for $x < x^\star$ and strictly negative for $x > x^\star$. This implies that $g'(x)$ is strictly decreasing for $x < x^\star$ and strictly increasing for $x > x^\star$. Since $g$ is strictly decreasing this shows that the magnitude of $g'$ is maximized at $x^\star$. ∎

Combining Lemma 3.3 with Fact 2.7, we immediately get Theorem 3.1. Lemma 3.3 further implies that the constant $c$ in the Theorem is $|g'(x^\star)|$, where $x^\star$ is the unique fixed point of $g$.

## 4 Strong spatial mixing on the $d$-ary tree

In this final section we use the message defined in the previous section to prove our main result, Theorem 1.1. Along with Weitz's reduction stated in Theorem 2.8, this will immediately imply Corollary 1.2, the FPTAS for the anti-ferromagnetic Ising model with arbitrary fields. To derive Corollary 1.3 for general two-state spin systems, we will need the translation described in Appendix A. Both these latter proofs appear at the end of this section.

Recall that Theorem 1.1 asserts that weak spatial mixing implies strong spatial mixing. We already showed in Theorem 3.1 that in the uniqueness region, there is uniform contraction in the tree recurrence with uniform boundary conditions. However, in order to prove strong spatial mixing, we will need to handle non-uniform boundary conditions as well, in which case the one-parameter recurrence $g = f^\phi$ is no longer sufficient. We therefore consider the multi-parameter vectorized version $G$ of the function $g$. For $\vec{x} \in \mathbb{R}^d$, $G(\vec{x})$ is defined as

$$G(x_1, x_2, \ldots, x_d) = \phi\left(\frac{1}{1 + \lambda \prod_{i=1}^d h(\psi(x_i))}\right), \quad (8)$$

where, as before, $\psi = \phi^{-1}$. We claim that strong spatial mixing on the $d$-ary tree is implied whenever the function $G$ satisfies the following condition; a proof of this implication can be found in Appendix C.2.

**Definition 4.1 (Contractive spatial mixing).** *Given the parameters $\beta$ and $\lambda$ for the anti-ferromagnetic Ising model on a $d$-regular tree, contractive spatial mixing holds if there exists a constant $c < 1$ such that*

$$|G(\vec{x}) - G(\vec{y})| \leq c\|\vec{x} - \vec{y}\|_\infty,$$

*for the vectorized version $G$ of $g$ defined as above with respect to the message $\phi$.*

To establish this condition, we will rely on the following lemma.

**Lemma 4.2** *Let $\eta = \psi(G(\vec{x}))$. Let $\bar{\eta}$ be the unique solution of $\psi(g(\bar{\eta})) = \eta$. Then $\|\nabla (G(\vec{x}))\|_1 \leq \|\nabla (G(\bar{\eta}, \bar{\eta}, \ldots, \bar{\eta}))\|_1 = |g'(\eta)|$.*

*Proof.* Set $\alpha_i = \psi(x_i)$ for $i = 1, 2, \ldots d$. We then have

$$\eta = \frac{1}{1 + \lambda \prod_{i=1}^d h(\alpha_i)} = \frac{1}{1 + \lambda h(\psi(\bar{\eta}))^d}. \tag{9}$$

Recalling the definitions of the quantities $A$ and $D$ given just before Lemma 3.2, we can now write $\|\nabla (G(\vec{x}))\|_1$ as

$$\|\nabla (G(\vec{x}))\|_1 = \frac{d\eta(1-\eta)(1-\beta^2)}{\beta + A\eta(1-\eta)} \left(1 + (1-\beta^2) \sum_{i=1}^d \frac{\alpha_i(1-\alpha_i)}{\beta + (1-\beta)^2 \alpha_i(1-\alpha_i)}\right). \tag{10}$$

For notational convenience, we define the function $J(x) \triangleq \frac{x(1-x)}{\beta + (1-\beta)^2 x(1-x)}$. Note that maximizing the sum in (10) under the constraint (9) is the same as maximizing $\sum_{i=1}^d J(\alpha_i)$ under the constraint that $\prod_{i=1}^d h(\alpha_i) = \frac{1-\eta}{\lambda \eta}$. Since $h$ is positive and invertible, it is therefore sufficient to show that the function $K(x) \triangleq J(h^{-1}(e^x))$ is concave in order to show that all $\alpha_i$'s are equal at a maximum. We now show this by direct computation. After differentiating twice and simplifying, we have

$$K''(x) = -\frac{e^{-x}(1 + e^{2x})\beta}{(1-\beta^2)^2} < 0.$$

This shows that $K$ is concave. By the discussion above, it follows that the sum in equation (10) is maximized when all $\alpha_i$'s are equal. In conjunction with the condition that $\eta = \frac{1}{1 + \prod_{i=1}^d h(\alpha_i)}$, this shows that

$$\|\nabla (G(\vec{x}))\|_1 \leq \|\nabla (G(\bar{\eta}, \bar{\eta}, \ldots, \bar{\eta}))\|_1.$$

Note that for any $x$, $G(x, x, \ldots, x) = g(x)$, and therefore $\|\nabla (G(\bar{\eta}, \bar{\eta}, \ldots, \bar{\eta}))\|_1 = |g'(\bar{\eta})|$. ∎

Using Lemma 3.3 and the above lemma, we are now ready to prove our main technical result, Theorem 1.1, which says that weak spatial mixing implies strong spatial mixing for general anti-ferromagnetic Ising models.

*Proof of Theorem 1.1.* Consider a setting of parameters $\beta$ and $\lambda$ such that the $d$-ary tree has weak spatial mixing. Let $x^\star$ be the unique fixed point of the function $g$. We will use only the property that the fixed point satisfies the condition (6) of Theorem 2.5. By Theorem 3.1 we have $\|g'\|_\infty = c < 1$. By Lemma 4.2, this implies that for all $\vec{x}$ in the domain of the function $G$ defined in equation (8), $\|\nabla G(\vec{x})\|_1 \leq c$. Using the mean value theorem followed by Hölder's inequality, we then have

$$|G(\vec{x}) - G(\vec{y})| \leq c \|\vec{x} - \vec{y}\|_\infty,$$

for all vectors $\vec{x}$ and $\vec{y}$ in the domain of $G$, and thus contractive spatial mixing holds. As observed above, this implies strong spatial mixing. ∎

**Remark:** We can now justify our use of the term "uniqueness region" as described in the Remark following Theorem 2.5. Notice that in the proof of Theorem 1.1 above, we used only the fact that weak spatial mixing implies that $(\beta, \lambda)$ is in the "uniqueness region" as defined in the aforementioned Remark. Thus, we see that whenever $(\beta, \lambda)$ is in the uniqueness region, we have strong spatial mixing, and hence, in particular, uniqueness. As stated earlier, this provides a rigorous proof of the claim in [3] that the interior of the uniqueness region is equivalent to the condition (6).

Combining the above theorem with the general reduction of Weitz [12] stated in Theorem 2.8, we can now prove Corollary 1.2, which asserts the existence of an FPTAS for general anti-ferromagnetic Ising models on bounded-degree graphs up to the uniqueness threshold.

*Proof of Corollary 1.2.* As observed earlier, in order to obtain an FPTAS for the partition function of the associated spin system, it is sufficient to give an FPTAS for approximating the occupation probability $p_\rho$ of a vertex $\rho$, under an arbitrary fixing of spin values for an arbitrary subset of vertices. Given a vertex $\rho$ in a graph $G$ of maximum degree $(d+1)$, we start by constructing Weitz's self-avoiding walk (SAW) tree rooted at $\rho$. For non-leaf vertices (apart from $\rho$) in this tree which do not have $d$ children, we can create dummy children (so as to make the arity of the vertex $d$) all of which independently have occupation probabilities of $1/2$. It is easy to see that this does not change the output of the tree recurrence (equation (5)) at any vertex of the tree. As we saw in the proof of Theorem 1.1, we have strong spatial mixing on this SAW tree whenever $(\beta, \lambda)$ are in the uniqueness region of the $d$-ary tree. The corollary now follows using Weitz's reduction (Theorem 2.8). ∎

Finally, we will see how to use Lemmas 3.2 and 4.2 to prove Corollary 1.3, which extends the FPTAS to general two-state anti-ferromagnetic spin systems.

*Proof of Corollary 1.3.* Given a two-state spin system with parameters $\beta, \gamma$ and $\lambda$ on a graph $G$ of degree at most $d+1$, we can use the translation given in Appendix A to come up with an equivalent Ising model with edge potential $\beta' = \sqrt{\beta\gamma}$ and vertex-dependent potentials $\lambda_v = \lambda(\sqrt{\gamma/\beta})^{d_v}$. Now, as before, in order to estimate the occupation probability $p_\rho$ for a given vertex $\rho$, we construct Weitz's self-avoiding walk (SAW) tree rooted at $\rho$, and complete the degree of any non-leaf vertex (apart from $\rho$) in the tree which does not have $d$ children by attaching dummy children which are fixed to have occupation probability $\frac{1}{2}$. We now use the message $\phi$ constructed above for $d$-ary trees for the parameter $\beta'$. By the hypotheses of the corollary, the parameters $(\beta', \lambda_u)$ at each vertex $u$ of the SAW tree are in the uniqueness region of the $d$-ary tree. Since the message $\phi$ does not depend upon $\lambda_u$, Theorem 3.1 and Lemma 4.2 apply at each vertex $u$ of the tree. Thus, as in the proof of Theorem 1.1, we get contractive spatial mixing and, hence, strong spatial mixing on the SAW tree. Employing Weitz's reduction (Theorem 2.8), we have the first part of the corollary.

The claim that the class $\mathcal{G}$ in the corollary includes $(d+1)$-regular graphs when $\beta, \gamma$ and $\lambda$ are in the uniqueness region of $d$-ary tree follows by noticing that in this case the parameters $\lambda' = \lambda_v$ obtained by the translation are the same at each vertex $v$, and that $\beta'$ and $\lambda'$ are in the uniqueness region of the $d$-ary tree by the hypotheses of the corollary. Thus, we can complete the proof for this case in the same manner as in the proof of Corollary 1.2. ∎

**Acknowledgments:** We thank Prasad Tetali for providing a manuscript of [9]. We also thank Dror Weitz, Colin McQuillan and an anonymous reviewer for helpful comments.

# Appendix

## A  Translation between various descriptions of the Ising model

General two-state spin systems are usually described in terms of (symmetric) energy functions $Q(+,+)$, $Q(+,-) = Q(-,+)$ and $Q(-,-)$, and an odd vertex field $h(+) = -h(-) = h$. For a graph $G = (V, E)$, the partition function of the system is then $Z_2 = \sum_\sigma w_2(\sigma)$, where the sum is over all states of the system $\sigma : V \to \{+, -\}$ and $w_2(\sigma)$ is defined as

$$w_2(\sigma) \triangleq \exp\left(-\sum_{\{u,v\} \in E} Q(\sigma(u), \sigma(v)) - \sum_{v \in V} h(\sigma(v))\right).$$

This is in fact equivalent to our formulation of the system given in Section 1.1 (see equations (1) and (2)). To see this, define

$$\begin{aligned}\beta &= \exp\left(-Q(+,+) + Q(+,-)\right); \\ \gamma &= \exp\left(-Q(-,-) + Q(+,-)\right); \\ \lambda &= \exp(2h),\end{aligned}$$

which yields

$$w(\sigma) = w_2(\sigma) \exp\left(Q(+,-)|E| + h|V|\right)$$

for all $\sigma$, and, therefore,

$$Z = Z_2 \exp\left(Q(+,-)|E| + h|V|\right).$$

We call the above spin systems *soft constraint* systems if $\beta, \gamma$ and $\lambda$ are non-zero, or equivalently, if the energy functions and field are finite for all spin values. As we shall now see, every such soft constraint system can be represented in terms of the Ising model (this translation can also be found, e.g., in [5]). Consider a general two-state spin system with parameters $\beta, \gamma > 0$ and $\lambda$. Then the equivalent Ising model has edge activity

$$\beta' = \sqrt{\beta\gamma},$$

and a degree-dependent vertex activity given by

$$\lambda'_v = \lambda \left(\sqrt{\frac{\gamma}{\beta}}\right)^{d_v},$$

where $d_v$ denotes the degree of vertex $v$. Now, denote the weight of a configuration $\sigma$ in the Ising model just defined by $w^\star(\sigma)$ and its partition function by $Z^\star$. Then one calculates straightforwardly that

$$w(\sigma) = w^\star(\sigma) \left(\sqrt{\frac{\beta}{\gamma}}\right)^{|E|}$$

and hence

$$Z = Z^\star \left(\sqrt{\frac{\beta}{\gamma}}\right)^{|E|}.$$

Thus we have translated the original spin system with parameters $(\beta, \gamma, \lambda)$ into an Ising model with locally changing field. Note that on regular graphs the resulting field is in fact constant at all vertices. Furthermore, the Ising model is anti-ferromagnetic if and only if $\beta\gamma < 1$. This justifies our use of the term "anti-ferromagnetic" for general spin systems based on the value of $\beta\gamma$. We also observe that in the special case of $d$-regular trees, this implies that weak (strong) spatial mixing in the original spin system $(\beta, \gamma, \lambda)$ is equivalent to weak (strong) spatial mixing in the Ising model given by the translation. A little thought shows that since all vertices except the root have the same degree in the $d$-ary tree, the last observation holds also for $d$-ary trees.

# B   Weak spatial mixing does not imply strong spatial mixing for the ferromagnetic Ising model

We construct a counterexample as follows: given a degree $d \geq 3$, consider the infinite rooted $d$-ary tree, with the fixed boundary condition where each vertex in the tree has one of its children fixed to +. Notice that if the original parameters are $\beta \geq 1$ and $\lambda$, the effect of this fixed boundary condition can be simulated by changing the vertex field to $\frac{\lambda}{\beta}$. Therefore, strong spatial mixing on this subgraph of the $d$-ary tree with parameters $(\beta, \lambda)$ holds only if weak spatial mixing holds on the $(d-1)$-ary tree with parameters $(\beta, \frac{\lambda}{\beta})$. It is therefore sufficient to choose $\beta$ and $\lambda$ satisfying both the conditions

$$\log \lambda > \log \lambda_c(\beta, d), \text{ and} \tag{11}$$
$$0 < \log \lambda - \log \beta < \log \lambda_c(\beta, d-1) \tag{12}$$

in order to construct a counterexample. To see that such a choice of parameters is possible, we consider the exact form of $\lambda_c(\beta, d)$. Translating the results in [3, p. 250] to our notation using Appendix A, we have

$$\log \lambda_c(\beta, d) = (d-1)\log\beta - P(d) + Q(\beta, d),$$

where

$$P(d) \triangleq d\log d - (d-1)\log(d-1), \text{ and}$$
$$\lim_{\beta \to \infty} Q(\beta, d) = 0, \text{ for any fixed } d.$$

We note that $P(d)$ is an increasing function of $d$. Thus, the required conditions (11) and (12) become

$$\log \lambda > \log \beta, \tag{13}$$
$$\log \lambda > (d-1)\log \beta - P(d) + Q(\beta, d), \text{ and} \tag{14}$$
$$\log \lambda < (d-1)\log \beta - P(d-1) + Q(\beta, d). \tag{15}$$

For a fixed $d$, $Q(\beta, d)$ is $o_\beta(1)$, and hence inequality (14) implies inequality (13) for $\beta$ large enough and $d \geq 3$. Again, since $Q(\beta, d)$ is $o_\beta(1)$, and $P(d)$ is an increasing function, it is possible to find $\lambda$ satisfying both the inequalities (14) and (15) for $\beta$ large enough. Thus, for any $d \geq 3$, we can find $(\beta, \lambda)$ with $\beta > 1$ such that weak spatial mixing for the $d$-ary tree does not imply strong spatial mixing.

## C Contraction and spatial mixing

### C.1 Contraction and weak spatial mixing

We show in this section that a stepwise contraction in the recurrence for $\phi(p_v)$ implies weak spatial mixing. As before, we denote $f^\phi$ by $g$, and assume that for any $x$, $y$ in the range of $\phi$, it holds that

$$|g(x) - g(y)| \leq c|x - y|, \tag{16}$$

for some $c < 1$. To show that this implies weak spatial mixing, we consider boundary conditions $\sigma_1$ and $\sigma_2$ on a set $S$ whose distance from the root $\rho$ is $l$. Using the monotonicity of the tree recurrence $F$ (defined in equation (5)) in all its arguments, it can be verified that $|\phi(p_\rho(\sigma_1, S)) - \phi(p_\rho(\sigma_2, S))|$ is maximized when $S$ is the set of all leaves at distance $l$ from $\rho$ and $\sigma_1$ assigns all vertices in $S$ to $+$ and $\sigma_2$ assigns all vertices in $S$ to $-$. With this definition of $\sigma_1$ and $\sigma_2$, we notice that the tree recurrence for $\phi(p_v)$ outputs the same value for all vertices $v$ at the same distance from the root. For a vertex at distance $l - i$ from the root $\rho$, we denote by $q_{i,j}$ the quantity $\phi(p_v(\sigma_j, S))$, for $j \in \{1, 2\}$. Now, using condition (16), we have

$$|q_{i+1,1} - q_{i+1,2}| = |g(q_{i,1}) - g(q_{i,2})|$$
$$\leq c|q_{i,1} - q_{i,2}|.$$

Since both $\phi$ and $\phi^{-1}$ are continuously differentiable functions defined over compact sets, they are Lipschitz continuous, say with parameters $L_1$ and $L_2$ respectively. We therefore have weak spatial mixing, since

$$|p_\rho(\sigma_1, S) - p_\rho(\sigma_2, S)| = |\phi^{-1}(q_{l,1}) - \phi^{-1}(q_{l,2})|$$
$$\leq L_2 c^l |q_{0,1} - q_{0,2}|$$
$$\leq L_1 L_2 c^l.$$

### C.2 Contraction and strong spatial mixing

In this section, we show that contractive spatial mixing, as defined in Definition 4.1 implies strong spatial mixing. We again consider boundary conditions $\sigma_1$ and $\sigma_2$ on a set $S$ which differ only on a subset $T$ which is at distance $l$ from the root $\rho$. Again, since both $\phi$ and $\phi^{-1}$ are continuously differentiable functions defined over compact sets, they are Lipschitz continuous, say with parameters $L_1$ and $L_2$ respectively. We define the quantity $q_i$ as

$$q_i \triangleq \max_{v:\delta(\rho,v)=l-i} |\phi(p_v(\sigma_1, S)) - \phi(p_v(\sigma_2, S))|.$$

Notice that $q_0 \leq |\phi(1) - \phi(0)| \leq L_1$. Also, since $G$ is the tree recurrence for $\phi(p_v)$, contractive spatial mixing for $G$ implies $q_{i+1} \leq cq_i$ for $c < 1$. Thus, we get strong spatial mixing since

$$|p_\rho(\sigma_1, S) - p_\rho(\sigma_2, S)| \leq L_2 q_l \leq L_2 c^l q_0 \leq L_1 L_2 c^l.$$

# D Proof of Lemma 3.2

In this section, we prove Lemma 3.2. The proof involves a few somewhat lengthy derivative computations, which we isolate in the following lemma.

**Lemma D.1** *With the notation used in Lemma 3.2 above, we have*

$$\frac{\phi''(x)}{\phi'(x)} = \frac{A(2x-1)}{\beta + Ax(1-x)}; \tag{17}$$

$$\frac{h'(x)}{h(x)} = \frac{1-\beta^2}{\beta + (1-\beta)^2 x(1-x)}; \tag{18}$$

$$\frac{h''(x)}{h'(x)} = \frac{2(1-\beta)}{1-(1-\beta)x}; \tag{19}$$

$$f'(x) = -df(x)(1-f(x))\frac{h'(x)}{h(x)}; \tag{20}$$

$$\frac{f''(x)}{f'(x)} = \frac{f'(x)(1-2f(x))}{f(x)(1-f(x))} + \frac{h''(x)}{h'(x)} - \frac{h'(x)}{h(x)}. \tag{21}$$

*Proof (sketch).* Most of these identities are easily verified by direct computation. In proving equation (17), one needs to keep in mind the definition of the constant $D$. ∎

*Proof of Lemma 3.2.* To ease notation, we will suppress the dependence of the quantities $\eta$ and $\alpha$ on $x$. Using the chain rule, we have

$$g'(x) = \frac{\phi'(\eta)}{\phi'(\alpha)} f'(\alpha).$$

Here, we used the fact that since $\psi = \phi^{-1}$, $\psi'(x) = \frac{1}{\phi'(\psi(x))}$. After taking the logarithm, and noticing that the right hand side is more easily expressed as a function of $\alpha$ rather than of $x$, one can write the second derivative of $g$ as

$$\frac{1}{\psi'(x)} \frac{g''(x)}{g'(x)} = \frac{\phi''(\eta)}{\phi'(\eta)} \frac{d\eta}{d\alpha} - \frac{\phi''(\alpha)}{\phi'(\alpha)} + \frac{f''(\alpha)}{f'(\alpha)}. \tag{22}$$

We now consider each of the terms involved above. Recalling that $\eta = f(\alpha)$, and using equations (20) and (21) to expand the first and last terms in equation (22) above, we get

$$\frac{1}{\psi'(x)} \frac{g''(x)}{g'(x)} = T_1 - T_2, \tag{23}$$

where $T_1$ and $T_2$ are defined as

$$T_1 \triangleq \frac{h''(\alpha)}{h'(\alpha)} - \frac{h'(\alpha)}{h(\alpha)} - \frac{\phi''(\alpha)}{\phi'(\alpha)}, \text{ and}$$

$$T_2 \triangleq d\frac{h'(\alpha)}{h(\alpha)} \left[ \frac{\phi''(\eta)}{\phi'(\eta)} \eta(1-\eta) + 1 - 2\eta \right].$$

Notice that all terms containing $\eta$ are isolated in $T_2$. We now consider each of the terms separately. For $T_1$, we have

$$\frac{h''(\alpha)}{h'(\alpha)} - \frac{h'(\alpha)}{h(\alpha)} = \frac{2(1-\beta)}{1-(1-\beta)\alpha} - \frac{1-\beta^2}{\beta + (1-\beta)^2\alpha(1-\alpha)}$$

$$= \frac{(1-\beta^2)(2\alpha-1)}{\beta + (1-\beta)^2\alpha(1-\alpha)}.$$

Here, we used equations (19) and (18) in the first line. Now using equation (17), we have

$$T_1 = \frac{(2\alpha - 1)\left((1-\beta)^2[\beta + A\alpha(1-\alpha)] - A[\beta + (1-\beta)^2\alpha(1-\alpha)]\right)}{(\beta + (1-\beta)^2\alpha(1-\alpha))(\beta + A\alpha(1-\alpha))}$$

$$= \frac{\beta(2\alpha - 1)((1-\beta)^2 - A)}{(\beta + (1-\beta)^2\alpha(1-\alpha))(\beta + A\alpha(1-\alpha))}$$

$$= \frac{-d\beta(2\alpha - 1)}{(\beta + A\alpha(1-\alpha))} \frac{h'(\alpha)}{h(\alpha)}.$$

Here, we use $A = d(1-\beta^2) + (1-\beta)^2$, followed by equation (18) in the last line.

We now consider $T_2$. Again using equation (17), we have

$$T_2 = d\frac{h'(\alpha)}{h(\alpha)}\left[\frac{A(2\eta - 1)\eta(1-\eta)}{\beta + A\eta(1-\eta)} - (2\eta - 1)\right]$$

$$= \frac{-d\beta(2\eta - 1)}{\beta + A\eta(1-\eta)} \frac{h'(\alpha)}{h(\alpha)}.$$

Notice that modulo the $\frac{h'(\alpha)}{h(\alpha)}$ factor, $T_1$ and $T_2$ have the same functional form as functions of $\alpha$ and $\eta$ respectively. In fact, the message $\phi$ is designed so as to make this possible. We can now substitute these values into equation (23) to get

$$g''(x) = d\beta g'(x)\psi'(x)\frac{h'(\alpha)}{h(\alpha)}\left[\frac{2\eta - 1}{\beta + A\eta(1-\eta)} - \frac{2\alpha - 1}{\beta + A\alpha(1-\alpha)}\right]$$

$$= d\beta g'(x)\psi'(x)\frac{h'(\alpha)}{h(\alpha)}\frac{(\eta - \alpha)(2\beta + A(\alpha\eta + (1-\alpha)(1-\eta)))}{(\beta + A\alpha(1-\alpha))(\beta + A\eta(1-\eta))}$$

$$= (\eta - \alpha)g'(x)\psi'(x)\frac{d\beta(1-\beta^2)(2\beta + A\alpha\eta + A(1-\alpha)(1-\eta))}{(\beta + \alpha(1-\alpha)(1-\beta)^2)(\beta + A\eta(1-\eta))(\beta + A\alpha(1-\alpha))},$$

where in the last step we used equation (18). ∎